\documentclass[9pt,twocolumn,twoside]{pnas-new}

\usepackage{amsmath,color,epsfig,setspace}

\templatetype{pnasresearcharticle}

\title{Old and new puzzles of gold tellurides: incommensurate crystal structure of calaverite AuTe$_2$ and predicted stability of AuTe.}

\author[a,b]{Sergey V.~Streltsov}
\author[c]{Valerii V. Roizen}
\author[a]{Alexey V. Ushakov}
\author[c,d]{Artem R. Oganov}
\author[e]{Daniel I.~Khomskii}

\affil[a]{Department of theory low-dimensional spin systems, Institute of Metal Physics, S. Kovalevskoy St. 18, 620990 Yekaterinburg, Russia}
\affil[b]{Department of theoretical physics and applied mathematics, Ural Federal University, Mira St. 19, 620002 Yekaterinburg, Russia}
\affil[c] {Computational materials discovery laboratory, Moscow Institute of Physics and Technology,  Institutskiy per. 9, 141701 Dolgoprudny, Moscow Region, Russian Federation}
\affil[d] {Materials Discovery Laboratory, Skolkovo Institute of Science and Technology, Nobel St. 3, 143026, Skolkovo, Russian Federation}
\affil[e]{II. Physikalisches Institut, Universit\"at zu K\"oln, Z\" ulpicher Stra\ss e 77, D-50937 K$\ddot o$ln, Germany}

\leadauthor{Lead author Streltsov} 

\significancestatement{
It is shown that the long-standing puzzle of incommensurate crystal structure of AuTe$_2$ can be solved, if this material is considered as a negative charge-transfer system. Using modern computational methods we demonstrate that charge redistribution associated with incommensurate modulations of crystal structure occurs not so much on Au, but predominantly on Te sites. This substantially reduces Coulomb energy costs for creating such a unique crystal structure. The same mechanism also explains superconductivity of doped AuTe$_2$. Exploring different  Au-Te compositions we also discovered a previously unknown compound AuTe, which theoretically is very stable, and we predict its crystal structure.}

\authorcontributions{The project was conceived by D.Kh. and S.S. The DFT calculations were performed by S.S. with help of A.U., V.R. and A.O. did study of all possible crystal structures for whole Au-Te series using USPEX. S.S. and D.Kh. analysed the results and wrote the manuscript with help of V.R. and A.O.}
\authordeclaration{The authors declare no competing interest.}
\correspondingauthor{\textsuperscript{2}E-mail: streltsov@imp.uran.ru}


\begin{abstract}
Gold is a very inert element, which forms relatively few compounds. Among them is a unique material - mineral calaverite, AuTe$_2$. Besides being the only compound in nature from which one can extract gold on an industrial scale, it is a rare example of a natural mineral with incommensurate crystal structure. Moreover, it is one of few systems based of Au, which become superconducting (at elevated pressure or doped by Pd and Pt). Using {\it ab initio} calculations we theoretically explain these unusual phenomena in the picture of negative charge transfer energy and self-doping, with holes being largely in the Te $5p$ bands. This scenario naturally explains incommensurate crystal structure of AuTe$_2$  and it also suggests a possible mechanism of superconductivity. {\it Ab initio} evolutionary search for stable compounds in the Au-Te system confirms stability of AuTe$_2 $ and AuTe$_3$ and leads to a prediction of a new stable compound AuTe, which until now has not been synthesized. 
\end{abstract}

\dates{This manuscript was compiled on \today}
\doi{\url{www.pnas.org/cgi/doi/10.1073/pnas.XXXXXXXXXX}}
\begin{document}

\verticaladjustment{-2pt}

\maketitle
\thispagestyle{firststyle}
\ifthenelse{\boolean{shortarticle}}{\ifthenelse{\boolean{singlecolumn}}{\abscontentformatted}{\abscontent}}{}

\section{Introduction}
 \dropcap{I}t is very well known that gold is one of the least reactive chemical elements and it is typically mined as a pure native element. It also occurs in alloys 
 but very rarely it can be found in the form of compounds. The only compound existing in nature from which one can extract gold on an industrial scale is gold telluride - AuTe$_2$, calaverite. This material is extremely interesting in many aspects. It even influenced the gold rush in Australia, where miners in gold mines first discarded calaverite as an “empty” waste and used it for paving the roads, but, after discovering that it contains real gold which can be extracted, very carefully scrapped all these roads.  

Another, very specific feature of AuTe$_2$ is that it is one of very few materials having in natural form an incommensurate crystal structure. This at a time gave a lot of headache to mineralogists and crystallographers: they could not understand peculiar faceting of calaverite crystals contradicting Ha\"uy's law. Usually the stable natural facets of a crystal are those with small Miller indexes, and in calaverite everything looked odd, until one realised that the very crystal structure is incommensurate \cite{Dam1985}. But the origin of the incommensurability is still obscure. Last but not least, AuTe$_2$ was found to be a superconductor at a relatively low pressure of 2.3 GPa or upon Pt or Pd  doping\cite{Kitagawa2013,Kudo2013,Ootsuki2014a,Reithmayer1993}, with critical temperature $\sim 4$~K.

In the present paper we found that all these properties of AuTe$_2$ can be naturally explained, if one takes into account that it is in a negative charge transfer energy regime, which drives a charge disproportionation resulting in an incommensurate crystal structure at normal conditions or superconducting state at higher pressures. Moreover, an extensive structural study of different gold tellurides allowed us to predict the existence  of a novel stable compound - AuTe. 
We report the predicted crystal structure and properties of this new material.

\section{Old puzzle of calaverite's crystal structure}

AuTe$_2$ has a distorted layered  CdI$_2$-type structure (the average structure has space group $C2/m$ \cite{Tunell1952}), with triangular layers of Au with Te atoms in between.   However, there is a periodic displacive modulation along [010] direction, which makes overall crystal structure incommensurate \cite{Tendeloo1983}. The mechanism of incommensurability is unclear. One may argue that it can be due a specific electronic structure, which results in a charge density wave (CDW) instability, but accurate band structure calculations have not found nesting of the Fermi surface at corresponding wave vectors\cite{Krutzen1999,Gonze2000}. Schutte and de Boer proposed another explanation based on the formal assignment of valencies in Au$^{2+}$(Te$_2$)$^{2-}$ (in analogy with another mineral – the “fool’s gold”  Fe$^{2+}$(S$_2$)$^{2-}$) \cite{Schutte1988}. However, whereas Fe$^{2+}$ is a stable ionic state, every chemist knows that Au$^{2+}$ is extremely difficult to stabilise: it exists as Au$^{1+}$($d^{10}$) or Au$^{3+}$ (nominally low-spin $d^8$). If one would manage to really stabilize Au$^{2+}$($d^9$), it would be a realization of an old dream - a ``magnetic gold''\footnote{It was actually indeed made, however not in oxides, but in systems with more ionic bonds – in Au(AuF$_4$)$_2$ and  Au(SbF$_6$)$_2$\cite{Elder1997}.}.

The phenomenon of skipped valence \cite{Varma1988} of Au$^{2+}$ can lead to the possibility of  charge disproportionation into Au$^{1+}$ and Au$^{3+}$, and it seems to naturally explain the ground state properties of AuTe$_2$, as it works for example in  Cs$_2$Au$_2$Cl$_6$ \cite{Ushakov2011}. The fact that CDW due to such charge disproportionation is  incommensurate in AuTe$_2$, in contrast to Cs$_2$Au$_2$Cl$_6$,  may be related to the triangular lattice, which  Au ions form in AuTe$_2$. This lattice is not bipartite, and the resulting frustration can lead to incommensurate modulation. While overall modulation of the lattice is complex, the local distortions seems to confirm this skipped valence interpretation: some Au ions, say at the maximum of CDW, are in a linear, or dumbbell coordination (two short and four long Au-Te bonds), typical for $d^{10}$ ions, here Au$^{1+}$, whereas at the ``other end'', say in the minimum of CDW, Au ions are square-coordinated – coordination typical for Au$^{3+}$($t_{2g}^6(3z^2-r^2)^2(x^2-y^2)^0$)\cite{Reithmayer1993}. Local surrounding for other Au interpolates between these two limits.

This interpretation, however, was put in doubt. First, photoemission\cite{VanTriest1990} and then x-ray absorption\cite{Ettema1994} measurements showed that apparently electronic configuration of all Au ions is the same, close to Au$^{1+}$.\footnote{Recent, spectroscopic study however did show the existence of slightly inequivalent Au ions\cite{Ootsuki2014a}.}  Also {\it ab initio} calculations performed for the artificial supercell structure with four Au ions mimicking the small-period CDW\footnote{Note that the structure used in Ref. \cite{Krutzen1999} is somehow unnatural in a sense that four short Au-Te bonds do not lie in one plane.} does not see any difference in occupation of $d-$shell for different Au ions\cite{Krutzen1999}.

We argue that nevertheless the physics of AuTe$_2$ is related to the eventual instability of Au$^{2+}$ against charge disproportionation, which determines main properties of AuTe$_2$, including not only incommensurate CDW, but also the tendency to superconductivity. As demonstrated below, the resolution of the controversy mentioned above lies in the fact that actually AuTe$_2$ is a negative charge transfer (CT) energy system\cite{Khomskii-97,Sawatzky2016}, with all the holes predominantly in the $5p$ bands of Te.

The notion of CT insulators was introduced in the seminal paper by Zaanen, Sawatzky, and Allen\cite{ZSA}. These are materials with strongly correlated electrons. However, the lowest charge excitations in them correspond not to transfer of electrons between localized $d$ states, $d^n d^n \to d^{n+1}d^{n-1}$, as in Mott-Hubbard insulators, but to electron transfer between TM’s and ligands, i.e. to the processes $d^n p^6 \to d^{n+1}p^5 = d^{n+1}\underline{L}$,  where $\underline{L}$ stands for the ligand hole. In CT insulators this CT excitation energy is positive, $\Delta_{CT} = E(d^{n+1}p^5) - E(d^np^6) > 0$,
but in principle it can be very small or even negative (naively speaking, when anion $p$ levels lie above $d$ levels of TM ions).  In this case we speak about negative CT energy. Usually this situation is met when the oxidation state of a metal is unusually high - e.g. 4+ for Fe or 3+ for Cu. If such states are created by doping, as, e.g., in high-T$_c$ cuprates the doped holes go predominantly to oxygen $p$ states (although these are of course strongly hybridized with $d$ states of Cu). But this situation can be realized also in undoped stoichiometric compounds such as CaFeO$_3$. In this case there can occur spontaneous transfer of electrons from ligands to TM ion, i.e. Fe$^{4+} \to $Fe$^{3+} \underline{L}$. This situation can be called a self-doping\cite{Korotin1998}. This picture, which in physics we describe by the (negative) CT energy, in chemistry is such notions as redox reaction, dative bonding etc. \footnote{A   ``dictionary'' helping to establish the correspondence between physical and chemical language is contained in the famous book by Goodenough\cite{Goodenough}, and a very clear paper by  Hoffmann\cite{Hoffmann2016a}.}

Interestingly enough, in many systems of this class there occurs spontaneous charge disproportionation, like 2Fe$^{4+} \to $Fe$^{3+} + $Fe$^{5+}$, but occurring predominantly on ligands, i.e. this ``reaction'' should be rather visualised as 2Fe$^{3+}\underline{L} \to $Fe$^{3+} + $Fe$^{3+} \underline {L}^2$. This process is now well established also in nickelates RNiO$_3$ (R=Pr, Nd), where it leads to a real phase transition, originally interpreted as charge ordering on Ni (2Ni$^{3+} \to$ Ni$^{2+}$ + Ni$^{4+}$)\cite{Alonso1999b}, but which is actually much better described by the ``reaction'' 2Ni$^{3+} \to$ Ni$^{2+}$ + Ni$^{1+}\underline{L}^2$ \cite{Mizokawa2000}.

We claim that the same phenomenon also occurs in systems containing Au$^{2+}$, such as, e.g., Cs$_2$Au$_2$Cl$_6$\cite{Ushakov2011}, and also in calaverite AuTe$_2$, where one can write this reaction as
\begin{eqnarray}
\label{reaction1}
2\mathrm{Au}^{2+} (d^9) \to \mathrm{Au}^{1+}(d^{10}) +\mathrm{Au}^{3+}(d^8), 
\end{eqnarray}
but in fact it should be rather visualized as 
\begin{eqnarray}
\label{reaction2}
2\mathrm{Au}^{2+} \equiv       2\mathrm{Au}^{1+} \underline{L} \to \mathrm{Au}^{1+} +\mathrm{Au}^{1+}\underline{L}^2.             
\end{eqnarray}
Two holes  ($\underline{L}^2$) in the Te $p$ band form something like a bound state, with the symmetry of a low-spin $d^8$ state of Au$^{3+}$ with which it hybridises. Below we confirm this picture by the {\it ab initio} band structure calculations.
\begin{figure}[t!] \centering
\includegraphics[width=1\columnwidth]{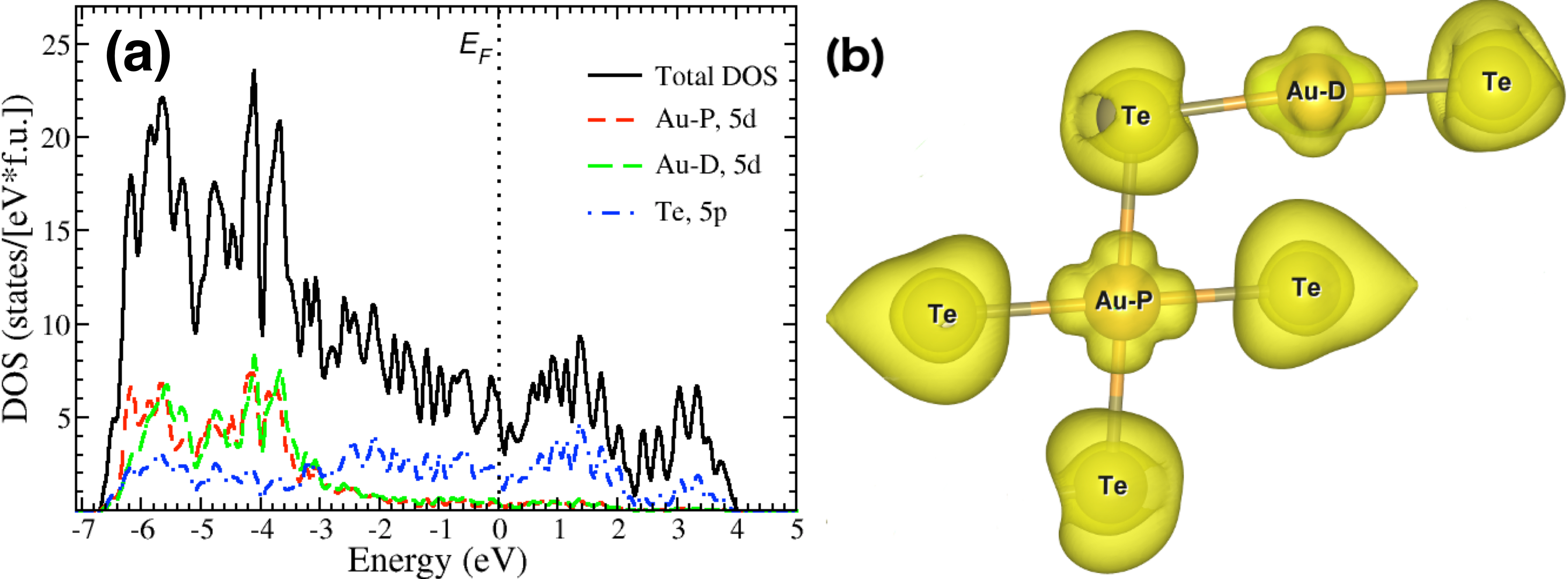}
\caption{(a) GGA+SOC total and partial density of states for AuAu'Te$_4$ structure (for experimental volume), (b) charge density $\rho(\vec r)$ corresponding to the top-most, partially filled band (isosurface corresponding to 0.003 $e/\AA^3$ $\approx$10\% of maximal value of $\rho(\vec r)$ is presented; the charge density was plotted using the PAW formalism). Results of the GGA+SOC calculations for AuTe$_2$ in fully optimized ``AuAgTe$_4$'' structure. Au-P and Au-D stands for Au ions having plaquette and dumbbell surroundings.}
 \label{charge-density}
 \end{figure}

\section{Mechanism of incommensurability in AuTe$_2$: Negative charge transfer energy}
 It is impossible to carry out {\it ab initio} calculations for the real incommensurate structure with the existing codes based on density function theory (DFT). One needs to approximate this structure by some supercell with a commensurate CDW. We borrowed an idea how to construct it from nature, taking initial crystal structure from the mineral sylvanite - AuAgTe$_4$.  Au and Ag ions in sylvanite are ordered in stripes, Ag being in a linearly-coordinated sites, typical for ions with $d^{10}$ configuration, i.e. it is Ag$^{1+}$ (Au-D in Fig.~\ref{charge-density}b), and Au ions lie in a square-coordinated positions, corresponding to, nominally, Au$^{3+}$ (Au-P in Fig.~\ref{charge-density}b) with its strong Jahn-Teller distortion  \cite{Tunell1952}. This structure was relaxed in the generalized gradient approximation (GGA) taking into account the spin-orbit coupling (SOC), then Ag was substituted by Au ion and relaxed again keeping unit cell volume the same as in real AuTe$_2$ (this structure will be labeled as AuAu’Te$_4$ in what follows).

First of all, we found that the AuAu’Te$_4$ structure is stable and there are still two differently coordinated Au.  Secondly, this structure is lower in energy than the average $C2/m$ structure \cite{Tunell1952} at experimental volume (by 22 meV/f.u.). Thus, one may see that we gain a lot of energy by making distortions corresponding to the CDW  (in this case a commensurate one). 

A close inspection of the Au $5d$ occupation numbers in the AuAu’Te$_4$ structure, however, show that from the point of view of $d-$occupation both Au ions are 1+: corresponding occupancies of the $d$ shell (as obtained within the projector augmented wave (PAW) method) are 9.90 and 9.92, so that the difference is negligible: $\delta n_{Au-d} = 0.02$ electrons (the Bader analysis\cite{bader1994} gives even smaller difference, $<$0.01 electron), while in real sylvanite, AuAgTe$_4$, $\delta n \sim 0.5$ electrons, i.e. in silvanite we can indeed speak about Ag$^{1+}$ and Au$^{3+}$  (Au$^{3+}$ again with a lot of ligand holes). This however seems to be in strong contrast with results of the lattice optimization, which  gives very different local coordination 
for two Au:  we have one linearly- (1+)  and another square-coordinated (3+) Au. The difference between short and long Au-Te bonds is $\sim$ 0.25\AA~in linear- and $\sim$ 0.35\AA~in square-coordinated Au. This is of order of magnitude of Jahn-Teller (JT) distortions in such classical JT systems as LaMnO$_3$ (0.27 \AA)\cite{LaMnO3-JT} and KCuF$_3$ ($\sim$0.3 \AA)\cite{Haegele1974}. What drives such strong lattice distortions, if not the CDW on Au sites?

In order to answer this question we plot in Fig.~\ref{charge-density}b the distribution of the charge density corresponding to the top-most, partially filled bands illustrating a hole distribution. One may see that there is only a minor contribution from the Au $5d$ states to the charge density corresponding to the least filled band, while the largest part comes from the Te $5p$ orbitals. Thus, one may speak about significant contribution of the ligand holes to the ground state wave function. The symmetry of ($\underline{L}^2$) hole state around ``Au$^{3+}$'' (Au-P in Fig.~\ref{charge-density}b) is the same as that of a hypothetical JT active Au$^{3+}$ ion with two holes on the $x^2-y^2$ orbital\footnote{Note that this orbital lies in the plane of Te plaquette, while central part of the charge density at Au-D is spherically symmetric and, thus, this band corresponds rather to $3z^2-r^2$ orbital.}, i.e. it naturally explains why this ion has square coordination typical for such a state. 

Analysis of the DOS, shown in Fig.~\ref{charge-density}a, also confirms that the largest part of the holes are in the Te $5p$ bands and one may speak about negative CT energy situation. The local electronic structure of Au ions in this case corresponds to 1+  valence state for all Au ions ($d^{10}$).  These results allow us to reconcile the picture of charge disproportionation driven largely by skipped valence of Au$^{2+}$, with the experimental  data \cite{VanTriest1990,Ettema1994}, which show that all Au ions are Au$^{1+}$ from spectroscopic point of view. 
\begin{figure}[t!] 
\centering
\includegraphics[width=1\columnwidth]{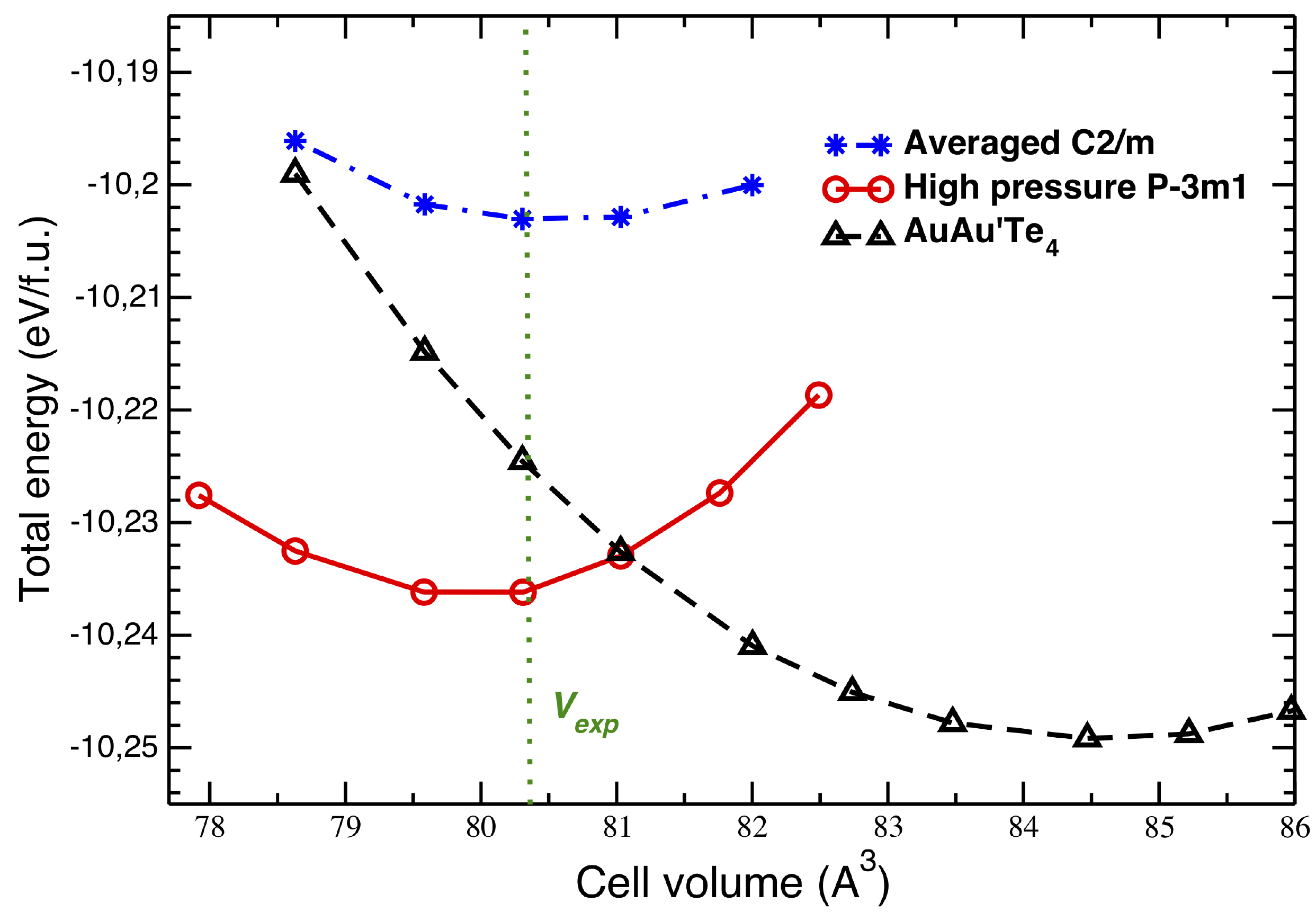}
\caption{Total energy  vs. volume for different possible crystal structures (GGA+SOC results). For each volume, optimization of the crystal structure was performed. }
 \label{energy-volume}
 \end{figure}

Moreover, a redistribution of electrons between Te and Au favors strongly distorted calaverite crystal structure, reminiscent of the formation of the CDW. Indeed, if the CT energy would be positive and there would be a real CDW with the Au$^{1+}$ and Au$^{3+}$ ions having $5d^{10}$ and $5d^8$ electronic configurations,  this would cost a lot of Coulomb energy (two holes on the same $d$ site repel each other with the energy $U$, which is $\sim$10 eV). Redistributing a part of the charge density to ligands we minimize the energy costs on the formation of the CDW. 

However, analysis of only two structures can be only qualitative. There is no guarantee, that there is no other structures, which would give a lower total energy. In addition the equilibrium volume in the DFT can be different from the experiment. In order to overcome the first difficulty we used the USPEX algorithm\cite{Oganov2006} to search for all possible structures of AuTe$_2$ with all experimentally known structures included in calculation. USPEX was previously successfully applied for investigation of structural properties of many different materials including those based on heavy metals\cite{Dong2017,Zhang1502,Zhang2013}. For AuTe$_2$ we found that the AuAu’Te$_4$ structure with distortions reminding the CDW still has the lowest total energy among hundreds of other structures obtained with the USPEX. There appear only two structures (in the interval of 100 meV/atom), which may compete with it: the high pressure $P\bar 3m1$ phase, where the incommensurate superstructure disappears and all Au ions become structurally equivalent, and a structure characterized by the $Cmmm$ space group, total energy of which is by 90 meV/atom higher than the one of  AuAu’Te$_4$.

In the second step we carefully checked how total energies of these crystal structures depend on the volume, see Fig.~\ref{energy-volume}. The AuAu'Te$_4$ structure corresponding to the CDW is still the lowest one, while the equilibrium volume is slightly overestimated. The next one is $P\bar 3m1$ with 5.5\% smaller volume, the average $C2/m$ and $Cmmm$ structures are much higher. 

At this stage one can demonstrate a crucial role of the CT energy for the formation of  the AuAu'Te$_4$ structure with distortions, imitating the real structure of AuTe$_2$. For this we performed model calculations, where the Au $5d$ bands were artificially shifted up in energy, thus increasing the CT energy and reducing the contribution of Te holes. We found that the shift on only 1 eV is enough to destabilize AuAu’Te$_4$ structure, and it makes the high-pressure $P\bar 3m1$ phase with all Au ions structurally equivalent the lowest in energy: the total energy difference is $E_{P\bar 3m1} - E_{AuAu’Te_4} \approx$ -2 meV/f.u. In the real AuTe$_2$, modelled by AuAu’Te$_4$,  the Au $5d$ states lie below Te $5p$, see Fig.~\ref{charge-density}a, which corresponds to a negative CT energy $\Delta_{CT}$. Shifting the Au $5d$ orbitals up leads to a decrease of absolute value of $\Delta_{CT}$ or even can make it positive. Then the charge disproportionation would have been mostly on the Au sites, which leads to a drastic increase of the energy costs of the CDW due to Coulomb interaction, as explained above, and as a result AuAu’Te$_4$ structure with inequivalent Au's becomes much higher in energy.

An important question is why in real AuTe$_2$ the superstructure is incommensurate. As explained above, due to calculation limitations we had to model it by the closest commensurate structure of a silvanite, our AuAu'Te$_4$. To check for the possibility to get incommensurate structure we calculated phonon spectrum\cite{phonopy} of AuTe$_2$. We indeed found that when we start from the homogeneous high-pressure phase $P\bar 3m1$, some phonon frequencies became imaginary with the minimal frequency at incommensurate wave vectors ${\bf q} \approx 0.41 {\bf a }+ 0.5 {\bf c}$ (where $\bf a$ and $\bf c$ correspond to the $P\bar 3m1$ structure), see Fig. S1(b). Thus, the real instability of homogeneous structure would indeed lead to an incommensurate superstructure. 

Very significantly, when we shift $d-$levels up, as explained above, these imaginary phonon frequencies disappear. This once again proves that the negative CT energy and corresponding large contribution of ligand holes are crucial for the formation of the incommensurate structure of AuTe$_2$.

\begin{figure}[t!] 
\centering
\includegraphics[width=1\columnwidth]{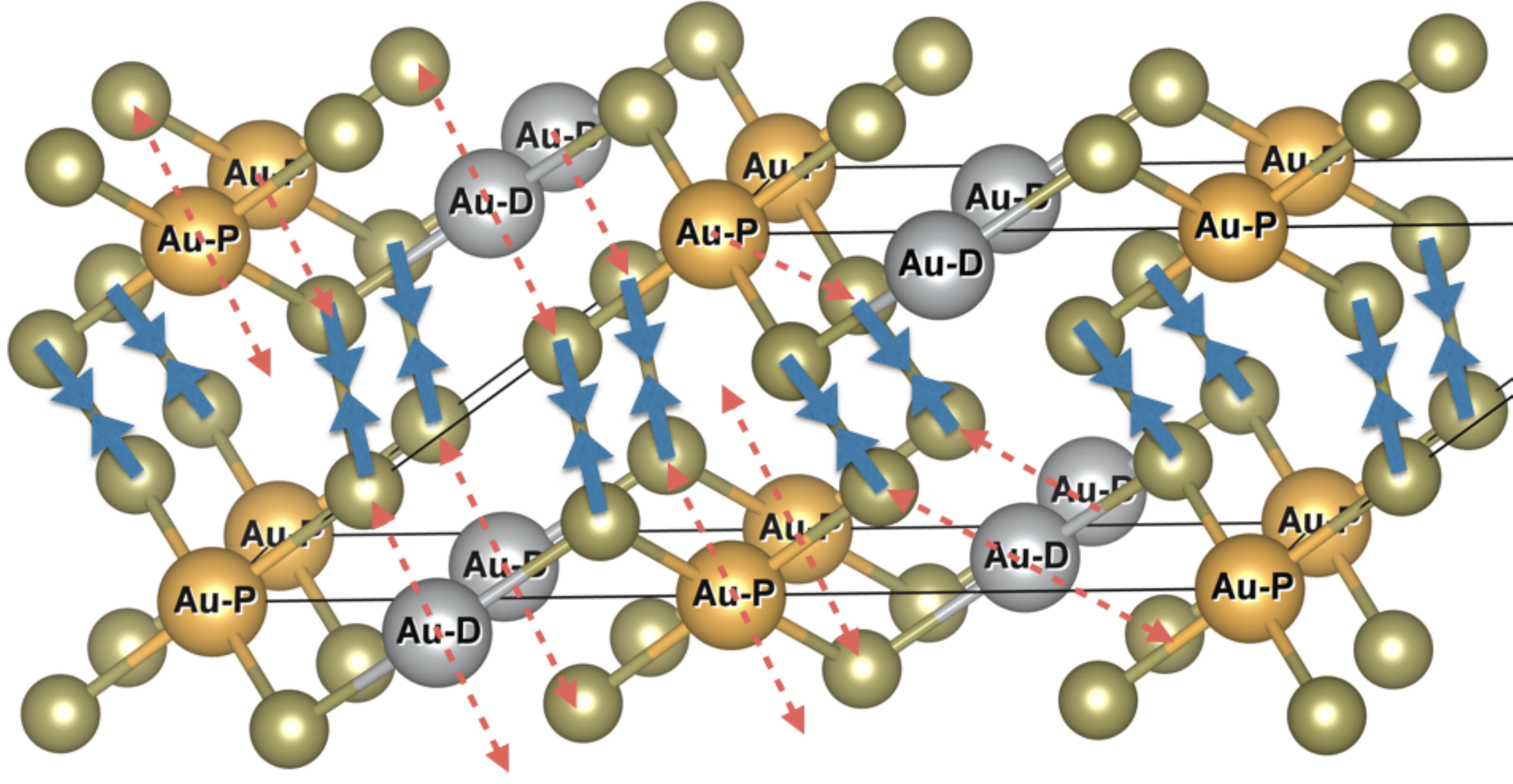}
\caption{Formation of the Te-Te dimers due to charge disproportionation on Au sites. 
The ``strength'' of distortions in AuTe$_6$ octahedra is not the same for all Au-Te bonds. There are ``strongly'' distorted with respect to undistorted $P\bar 3m1$ ($\delta_{Au-Te} \sim 0.45-0.55 $ \AA) and ``weakly'' distorted Au-Te bonds ($\delta_{Au-Te} \sim 0.15$\AA). Plotting (for simplicity) only ``strongly'' distorted Au-Te bonds (red lines; arrows show direction of distortions) one immediately obtains Te-Te dimers (shown by blue arrows).}
 \label{form-dimers}
 \end{figure}

\section{High pressure phase and superconductivity}
Taking first derivative of $E(V)$  one can find that a critical pressure ($P_c$) required for the transition from AuAu'Te$_4$ to $P\bar 3m1$ is 2.6 GPa. It is striking that while our optimized structure with the commensurate CDW (AuAu'Te$_4$) slightly overestimates equilibrium volume, the critical pressure for the transition to uniform $P\bar 3m1$ is reproduced with good accuracy: the  experimental  $P_c=2.5$ GPa~\cite{Reithmayer1993}.

The high-pressure phase of AuTe$_2$ is also very interesting due to another aspect - the superconductivity, which appears in it below $T_c=$2.3 K\cite{Kitagawa2013}. One may stabilize this phase not only by an pressure, but also by Pt doping\cite{Kudo2013}, which also results in the stabilization of the same $P\bar 3m1$ structure. The superconductivity was proposed to be induced by breaking of Te-Te dimers, which exist in the $C2/m$ phase, but disappear in the high-pressure superconducting $P\bar 3m1$ phase\cite{Kudo2013}. In particular it was speculated that the formation of Te-Te dimers  modifies electronic structure of AuTe$_2$ though formation of bonding ($\sigma$) and antibonding  ($\sigma^*$) Te $5p$ bands\cite{Kudo2013}. We have seen that the bands at the Fermi level indeed have very large contribution of the Te $5p$ states, but they are strongly hybridized with Au $5d$ and have the symmetry of Au $5d$ orbitals (see Fig.~\ref{charge-density}), while the $\sigma-$bonded Te $5p$ states are far away from the Fermi level ($\sim$5.2 eV below and $\sim$3.2 eV above $E_F$). Thus, it seems that the Te-Te dimerization is not directly related to the suppression of the superconductivity. In fact, this is just one of the consequences of the formation of the CDW. In Fig.~\ref{form-dimers} the directions of Te atoms displacements due to the CDW are indicated. One may see that the formation of AuTe$_4$ plaquettes and AuTe$_2$ dumbbells naturally results in dimerization of the Te atoms, which however is not a driving force but rather a consequence of the CDW formation in AuTe$_2$.

One can argue that the physics disclosed in our calculations, specifically the origin of the incommensurability, - the tendency to the skipped valence and charge disproportionation of ``Au$^{2+}$'', occurring in the situation with negative CT energy with the self-doping – is also instrumental in providing a mechanism of superconductivity in AuTe$_2$ under pressure or with doping. This tendency, both on the $d-$levels, reaction  \eqref{reaction1}, and more realistically on ligand states, reaction  \eqref{reaction2}, means that there exists a tendency for holes to form pairs, i.e. there exists an effective attraction of these holes. 

The idea that the tendency to charge disproportionation (which actually means the local ``chemical'' tendency to form pairs of electrons or holes) can be instrumental in providing the mechanism of Cooper pairing was first suggested by Rice and Sneddon\cite{Rice1981} in connection with the  superconductivity of doped BaBiO$_3$.  This material is also known to experience charge disproportionation of the type 2Bi$^{4+}\to$ Bi$^{3+}$ + Bi$^{5+}$ (and again with a lot of action on ligands - see, e.g., \cite{Sawatzky2016}). For high-$T_c$ cuprates similar idea was proposed in \cite{Khomskii1988}. It is also  closely related to some theoretical studies of superconductivity in systems with coexisting ordinary electrons and bipolarons, see e.g. \cite{Robaszkiewicz87,Friedberg1989}.  We suppose, by analogy with the above-mentioned papers, that the ``chemical'' tendency of Au$^{2+}$ to charge disproportionate into, nominally, Au$^{1+}$ and ``Au$^{3+}$'', which is the main ingredient of our theory  and which, as we argued above, plays crucial role in explaining the main properties of AuTe$_2$,  may be also instrumental in providing the mechanism, or at least helping the realization of superconductivity in AuTe$_2$ when doped or under pressure. 
\begin{figure}[t!] 
\centering
\includegraphics[width=0.7\columnwidth]{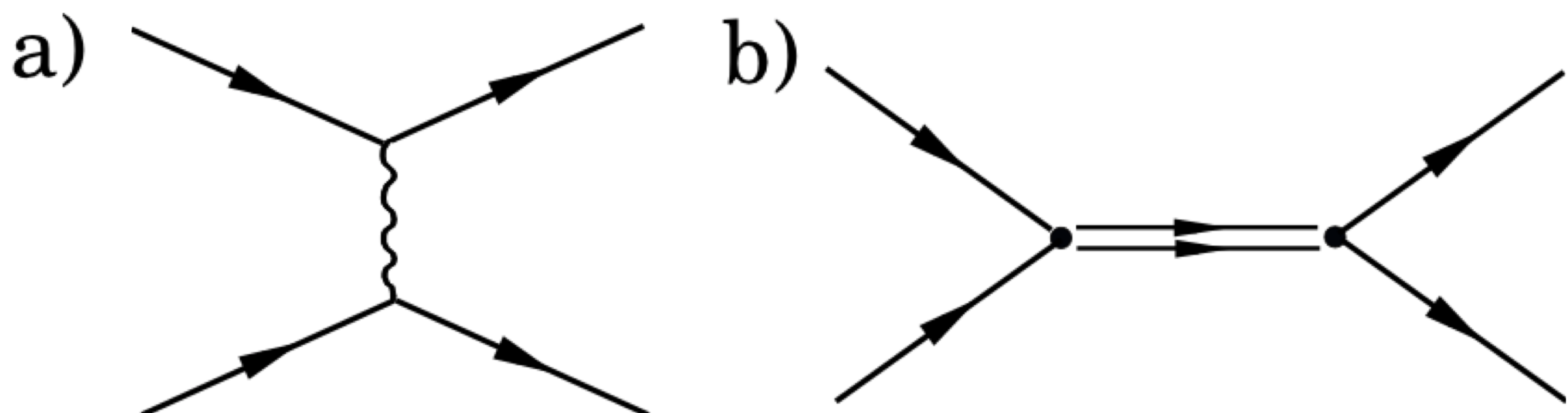}
\caption{Diagrams illustrating (a) conventional BCS ``t-''channel pairing and (b) ``s-''channel pairing proposed for AuTe$_2$.}
 \label{diagrams}
 \end{figure}

One can phenomenologically describe this situation by an effective Hamiltonian like the Anderson lattice model (where $5p$ electrons of Te play a role of conduction electrons, while $5d$ electrons of Au are localised), but with an effective attraction – with negative $U$  on localized levels. After excluding $d$ electrons, we get in effect also an attraction of conduction electrons, which, on one hand, can provide mechanism of CDW formation (not even requiring nesting of the Fermi-surface, although nesting would help). And, on the other hand, in this model we have a natural mechanism of formation of Cooper pairs leading to superconductivity. On the diagramatic language, this mechanism of pairing is described by the Fig.~\ref{diagrams}b (two electrons (or holes) of a conduction band ``drop'' into the Au $5d$ levels, where they experience attraction and form pairs, before decaying again into conduction electrons. This situation is reminiscent of a model with bipolarons\cite{micnas1990}, and is different from the usual electron-phonon exchange of Fig.~\ref{diagrams}a (although the standard electron-phonon coupling could also contribute).  Thus, AuTe$_2$ may be the long-sought second example of the same physics as proposed for BaBiO$_3$ \cite{Rice1981}, with the same mechanism of both the charge disproportionation and of the superconductivity.

\section{Novel compound: AuTe} 

\begin{figure}[t!]
\centering
\includegraphics[width=1\columnwidth]{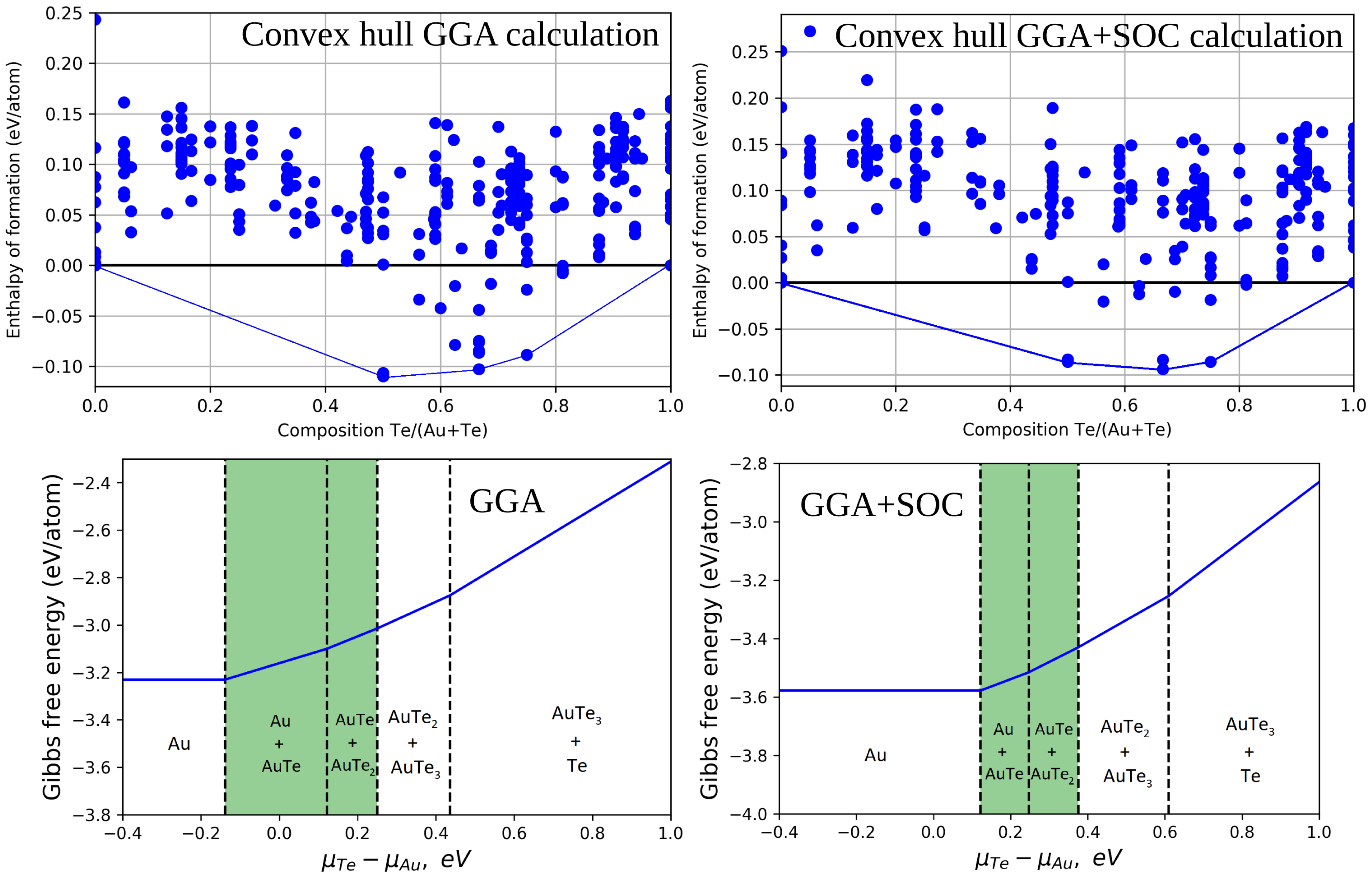}
  \caption{\label{Gibbs_mu}Thermodynamic convex hulls and Gibbs free energy G versus chemical potential $\mu$ for Au-Te system with different Te concentrations.}
\end{figure}

Since USPEX has shown its efficiency in determining the AuTe$_2$ crystal structure, we extended these calculations to a whole Au$_{1-x}$Te$_x$ series with arbitrary $x$. Fig.~\ref{Gibbs_mu}(a,b) shows thermodynamic convex hulls and phase diagram of the Au-Te system in the GGA and GGA+SOC approximations. A compound is thermodynamically stable if its thermodynamic potential (e.g., the Gibbs free energy) is lower than that of any other phase or phase assemblage of the same composition. On a graph showing the enthalpy of formation of all compounds of a given system (e.g. Au-Te system) from the elements, all points corresponding to stable compounds can be connected to form a convex hull. Height above the convex hull is a measure of thermodynamic instability of a compound. One may notice that in addition to experimentally observed structures as AuTe$_2$ and AuTe$_3$\cite{Luo1964} there appears a new one: AuTe. 

\begin{figure}[t]
 \begin{center}
  \includegraphics[height=0.5\linewidth]{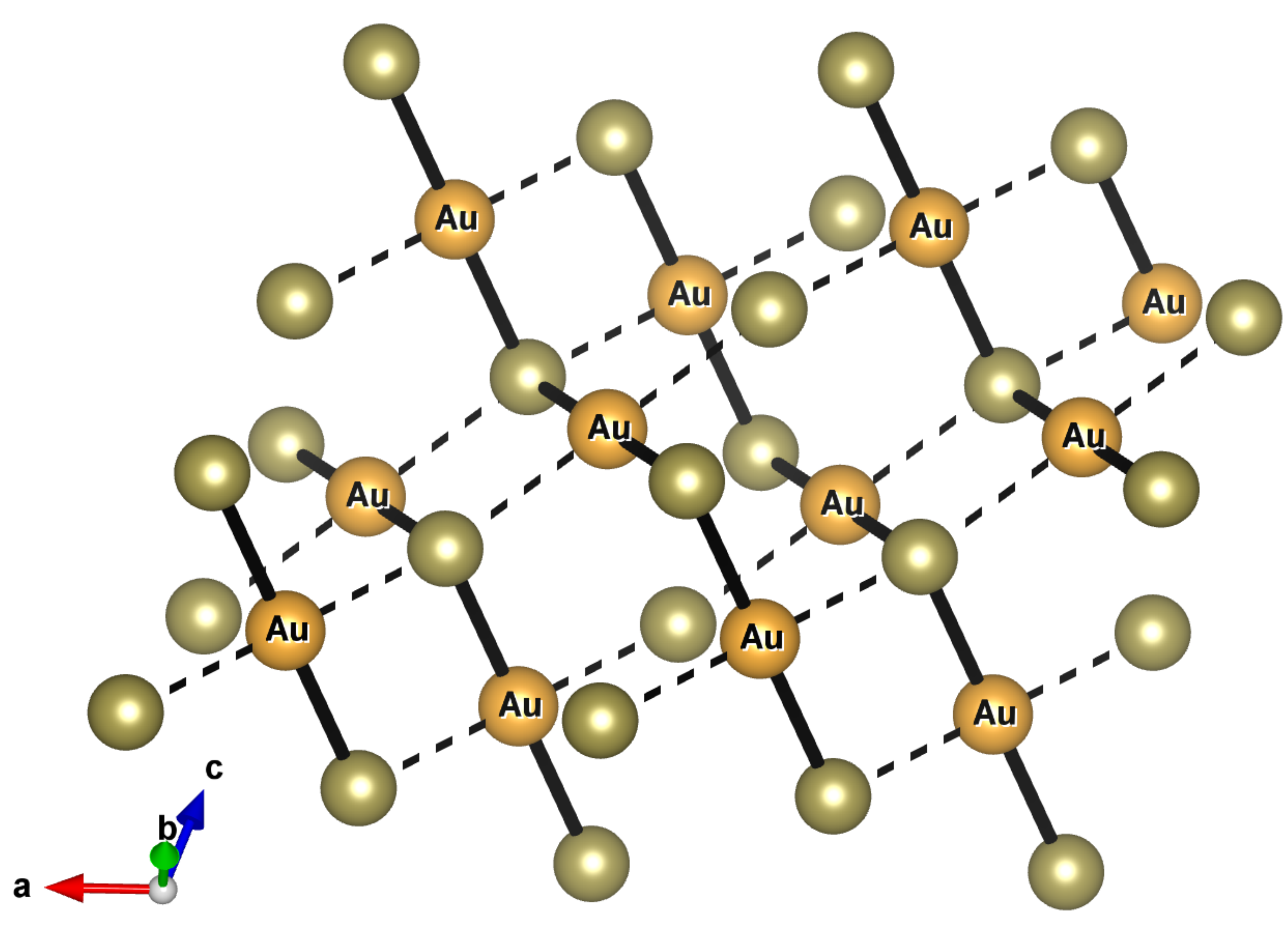} 
 \end{center}
\caption{  \label{AuTe-structure}The crystal structure of novel material: AuTe. Bold solid and dashed lines correspond to short and long Au-Te bonds, respectively.}
\end{figure}

AuTe has never been synthesized so far, but there exists mineral muthmannite, AuAgTe$_2$, found in Western Romania\cite{Bindi2004}, where Au and Ag ions are in 1:1 ratio. Muthmannite has a distorted NiAs-type structure with space group {\it P2/m}. Our calculations have shown that the {\it C2/c} structure predicted for AuTe by USPEX is significantly more stable (by 0.164 eV/atom with SOC) than muthmannite structure. 
The predicted {\it C2/c}  structure of AuTe, shown in Fig.~\ref{AuTe-structure}, can be considered as distorted NaCl-type structure (NiAs and NaCl structures are relatives). The Au ions are in the strongly distorted plaquettes with two short (2.68 \AA) and two long (2.90  \AA) Au-Te bonds.

It is worthwhile mentioning that the SOC additionally lowers position of the Au $5d$ band and thus affects stability of different phases in Au-Te system. One can see from Fig.~\ref{Gibbs_mu} that while both GGA and GGA+SOC calculations show stability of the same phases and crystal structures, there are large changes in stability fields. The plot of Gibbs free energy vs. chemical potential demonstrates that inclusion of the SOC expands stability field of Au (in effect making it more inert) and AuTe$_2$, at the expense of shrinking stability fields of AuTe and AuTe$_3$. The relatively narrow stability field may explain why AuTe is not yet known. 

AuTe was found to be a nonmagnetic metal in the GGA+SOC calculations. Analysis of the charge density, $\rho(\vec r)$, corresponding to the bands at the Fermi level, shows that there are nearly equal contributions to $\rho(\vec r)$ from Au $5d$ and Te $5p$ states. This may explain why USPEX did not find the solution corresponding to charge disproportionation, as it did for calaverite (two inequivalent Au ions: in dumb-bells and plaquettes): the energy costs due to the on-site Coulomb repulsion are too large in AuTe. Thus in effect AuTe should resemble the high-pressure phase of AuTe$_2$, with all Au equivalent, and one could expect that it could also be superconducting.
 
 \section{Conclusions}
The Au-Te system presents an interesting example of compounds of a very inert element, gold with nontrivial properties. We found out that there exist in the Au-Te system three stable stoichiometric compounds: AuTe, AuTe$_2$ and AuTe$_3$ \footnote{There exists also Au$_3$Te$_7$ with a simple cubic structure and statistical distribution of Au and Te atoms\cite{Newkirk1971} and is likely a solid solution. We have not found a stable compound with such stoichiometry in calculations at T=0 K, which indicates that it is probably entirely entropy-stabilized.}. The second and the less ``popular''  third compound are known and studied. AuTe has not been synthesized yet, although a similar material, mineral muthmannite AuAgTe$_2$, is known. It would be very interesting to check our predictions and try to synthesize and study AuTe. 

Much better studied, but still presenting several, until now unresolved puzzles, is calaverite AuTe$_2$. This is the system, the properties of which we now explained on the basis of {\it ab initio} calculations. The  picture emerging from our calculations is the following: The nominal average valence of gold in AuTe$_2$ is 2+, similar to many pyrites like FeS$_2$, MnS$_2$ etc. and layered dichalcogenides MS$_2$, MSe$_2$, MTe$_2$, where M - different TM ions\cite{Wilson1969}. But this state is, first of all, chemically unstable (only Au$^{1+}$ and Au$^{3+}$ are known to exist, with very few exceptions). And, most importantly, both ``Au$^{2+}$'' and ``Au$^{3+}$'' in AuTe$_2$ correspond to the situation with negative charge transfer energy, i.e.  practically Au$^{2+} \to  $Au$^{1+}\underline{L}$ and Au$^{3+} \to $Au$^{1+} \underline{L}^2$. This means that in fact all the holes go  to ligand (here Te) bands (but still with significant hybridization with $d$ states of Au). This is actually the situation of {\it self-doping}\cite{Korotin1998,Sawatzky2016}. In this case there occurs a phenomenon met also in several other systems: the valence, or charge disproportionation, which however again occurs not so much on the $d$ shells themselves, but on ligands. I.e. corresponding disproportionation is described not as  a \eqref{reaction1}, but rather as \eqref{reaction2}. This transition is accompanied (and is largely driven by) the change of the Au-Te bond lengths (and local coordination - linear for Au$^{1+}$ and square for ``Au$^{3+}$'' = Au$^{1+}\underline{L}^2$), i.e. it should be better called not charge, but bond  disproportionation \cite{Sawatzky2016}. But the outcome is very similar: there occurs in this case a structural transition with the formation of corresponding superstructures, commensurate as, e.g., in nickelates RNiO$_3$ \cite{Alonso1999b,Mizokawa2000,Sawatzky2016} or incommensurate as in the case of frustrated triangular lattice of AuTe$_2$. This picture naturally explain both the structural characteristics of AuTe$_2$ and the  spectroscopic data, showing apparently constant occupation of $d$ shells of Au. Despite this equivalence, the tendency to this charge, or bond disproportionation is intrinsically connected with the ``atomic'' property of, here, Au (skipped valence Au$^{2+}$). Suppression of this superstructure by pressure or doping leads to the formation of homogeneous metallic state with all Au (or Ni in RNiO$_3$) becoming equivalent, and in AuTe$_2$ this state becomes superconducting. The reverse charge transfer through ligand holes is a solid-state analogy of dative bonding known in coordination chemistry.

We argue that the same mechanism - the tendency to charge disproportionation, which is in fact the tendency to form electron or hole pairs, may be instrumental for the appearance of superconductivity in doped AuTe$_2$ or AuTe$_2$ under pressure. Thus, this exciting material, gold telluride, indeed is extremely interesting, both as to its rich history, but, more important for us, as an example of a very interesting physics.

\section{Methods}
The DFT calculations were performed within the Perdew-Burke-Ernzerhof  functional\cite{Perdew1996} using the all-electron PAW method\cite{Blochl1994} as realized in the VASP code\cite{Kresse1996}. We  took into account the SOC and used scalar-relativistic GW PAW potentials with an [Xe] core (radius 2.1 a.u.) and [Kr] core (radius 2.2 a.u.) for Au and Te atoms, respectively, and plane wave cut-off of 400 eV.  The evolutionary structure prediction algorithm USPEX\cite{Oganov2006} were applied for the search of different structural phases. Structure relaxations employed k-mesh with resolution of 2${\pi} \times 0.03\AA^{-1}$ and electronic smearing of 0.1 eV. The USPEX simulation included 80 structures per generation for variable-composition run. Also all known Au-Ag-Te compounds (with silver atoms substituted by gold) were included to the calculation\cite{Tunell1952,Newkirk1971,Luo1964,Belov1978,Bachechi1971}. Phonon calculations were performed using Phonopy\cite{phonopy} with $4 \times 4 \times 2$ supercell.

\section{Acknowledgements} 
Authors are grateful to G. Sawatzky, S.-W. Cheong, P. Becker, and L. Bohaty for discussions. This work was supported by the UB of RAS (18-10-2-37), by the RFBR (16-32-60070), by the FASO (``spin'' AAAA-A18-118020290104-2), and by Russian ministry of science and high education (02.A03.21.0006), by the DFG (SFB 1238), and by the German Excellence Initiative. A.O. thanks Russian Science Foundation  (16-13-10459), V.R. was supported by the Project 5-100 of MIPT, their computations were performed on the Rurik supercomputer.

\bibliography{library}

\end{document}